\documentclass[preprint2,twoside]{hwo}

\usepackage{graphicx}

\input{hwo.h}

\begin{document}

\title{\textbf{\LARGE Exploring the Quiescent Black Hole Population of Nearby \\ Dwarf Galaxies with the Habitable World Observatory}}
\author {\textbf{\large Fabio Pacucci$^{1,2}$}}
\affil{$^1$\small\it Center for Astrophysics $\vert$ Harvard \& Smithsonian, 60 Garden St, Cambridge, Massachusetts, USA; \\ $^2$\small\it Black Hole Initiative, Harvard University, 20 Garden St, Cambridge, Massachusetts, USA;\email{fabio.pacucci@cfa.harvard.edu}}


\author{\footnotesize{\bf Endorsed by:}
Torsten Boeker (European Space Agency),
Nico Cappelluti (University of Miami),
Filippo D'Ammando (INAF-IRA Bologna),
Ruben Joaquin Diaz (NSF NOIRLab),
Chris Impey (University of Arizona),
Stephanie LaMassa (Space Telescope Science Institute),
Valentina La Torre (Tufts University),
Eunjeong Lee (EisKosmos (CROASAEN), Inc.),
Athina Meli (North Carolina A\&T State University),
Andrea Sacchi (Center for Astrophysics $\vert$ Harvard \& Smithsonian,
Vivian U (Caltech/IPAC), and
Sylvain Veilleux (University of Maryland, College Park)
}



\begin{abstract}
  The formation and growth of supermassive black holes (SMBHs) remain a significant unsolved problem in astrophysics, particularly in the low-mass regime, where observations are sparse. The \textit{Habitable Worlds Observatory} (HWO), with its diffraction-limited imaging and high-resolution UV-optical spectroscopy, presents a unique opportunity to explore the demographics of massive black holes (MBHs) in nearby dwarf galaxies. We propose a program to dynamically detect quiescent MBHs with masses as low as $10^{4.5} \, \rm M_\odot$ in galaxies within $10-30$ Mpc, by resolving stellar velocity dispersions down to 30 $\rm km \, s^{-1}$. This effort will dramatically extend current black hole-host galaxy scaling relations into the dwarf regime, probing the fundamental connection between black hole seeds and their environments. Using a volume-limited sample of $\sim 100$ dwarf galaxies drawn from existing catalogs, HWO will resolve the gravitational sphere of influence of MBHs and enable precision measurements of stellar kinematics via Ca~II triplet absorption lines. These observations will test predictions from competing seed formation scenarios---light seeds from Population~III remnants versus heavy seeds from direct collapse---and clarify whether dwarf galaxies retain imprints of their initial black hole population. The results will offer critical insight into SMBH formation channel and the co-evolution of black holes and galaxies across cosmic time.
  \\
  \\
\end{abstract}

\vspace{2cm}

\section{Science Goal}

The fundamental question we aim to address in this HWO science case is: How did black holes and their host galaxies co-evolve to form the structures we observe today? 
This broad question can be split up into several questions, such as:
\begin{itemize}
    \item What is the role of supermassive black holes (SMBHs) in the formation and evolution of galaxies, from the early Universe to the local one?
    \item What fundamental mechanisms drive the co-evolution of SMBHs and their host galaxies?
    \item Did heavy black hole seeds form in the very distant Universe?
\end{itemize}

\subsection{Topics Related to the Astro2020:}
\begin{itemize}
    \item \textit{Black Hole Growth and Feedback:} Understanding the role of SMBHs in galaxy formation and evolution.
    \item \textit{Galaxy Formation and Evolution:} Investigating how galaxies and their central black holes have co-evolved from the early universe to the present day.
    \item \textit{Multi-Messenger Astronomy:} Combining electromagnetic observations with gravitational wave detections to study SMBH mergers and their impacts on galaxy evolution.
\end{itemize}

In particular, this Science Case is relevant to the following Key Science Questions and Discovery Areas of the Astro2020 Decadal Survey Report:

\begin{itemize}
    \item \textbf{B-Q2.} What Powers the Diversity of Explosive Phenomena Across the Electromagnetic Spectrum?
    \item \textbf{B-Q4.} What Seeds Supermassive Black Holes and How Do They Grow?
    \begin{itemize}
        \item \textbf{B-Q4a.} How Are the Seeds of Supermassive Black Holes Formed?
        \item \textbf{B-Q4b.} How Do Central Black Holes Grow?
    \end{itemize}
    \item \textbf{D-Q.} How did the intergalactic medium and the first sources of radiation evolve from cosmic dawn through the epoch of reionization?
    \begin{itemize}
        \item \textbf{D-Q3.} How do supermassive black holes form, and how is their growth coupled with the evolution of their host galaxies?
        \begin{itemize}
            \item \textbf{D-Q3a.} The seeds of supermassive black holes.
            \item \textbf{D-Q3b.} Existence and formation of inter\-med\-i\-ate-mass black holes.
            \item \textbf{D-Q3c.} Comprehensive census of supermassive black hole growth.
            \item \textbf{D-Q3d.} The physics of black hole feedback.
        \end{itemize}
    \end{itemize}
\end{itemize}

\subsection{Relevance for Other Broad Scientific Areas in the Astro2020 Decadal Survey:}
\begin{itemize}
    \item Panel on Cosmology (C)
    \item Panel on Particle Astrophysics and Gravitation (L)
\end{itemize}

\begin{figure*}[ht!]
    \centering
    \includegraphics[width=0.75\textwidth]{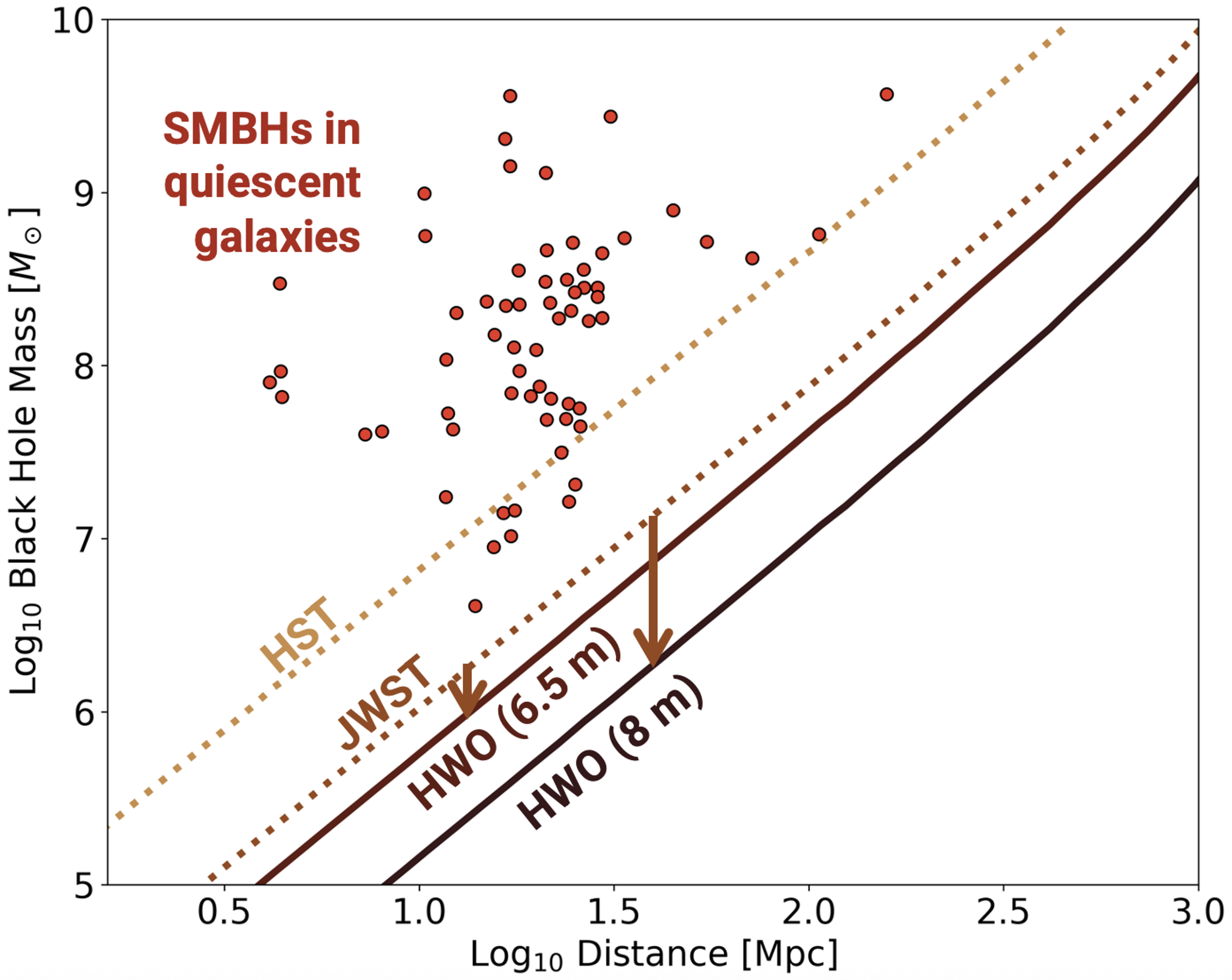}
    \caption{The minimum black hole mass that existing telescopes and HWO can dynamically detect as a function of the angular diameter distance. We assume that the dynamical mass is measured via standard optical absorption features (e.g., the Ca~II triplet lines). HWO in the 8-m (6.5-m) configuration can detect central MBHs up to 30 (10) Mpc. Compared to JWST, the smallest configuration can detect black holes that are $\sim$3 times lighter. In its extended configuration, HWO can detect black holes an order of magnitude lighter. The extended configuration also increases the explorable volume by a factor of $\sim$30.}
    \label{fig:fig1}
\end{figure*}

\section{Science Objective}

SMBHs are at the center of massive galaxies, and they have co-evolved over cosmic time, establishing a fundamental relationship between the mass of the black hole and the properties of the host galaxy \citep{Ferrarese_2000, Gebhardt_2000}. Studying this relation in smaller galaxies (thought to host lighter black holes) is beyond our current reach. HWO can be fundamental in understanding the co-evolution between black holes and galaxies, providing profound insights into the process that formed the first stars and black holes close to the beginning of time. 

\subsection{Primary Science Objective}

Dynamically detect the presence of quiescent (i.e., not accreting) massive black holes (MBHs) at the center of local (i.e., $< 10$--$30$ Mpc) dwarf galaxies via measuring the velocity dispersion of the stars at the core of the galaxy.

Due to its immense gravitational field, the stellar velocity dispersion is higher when a central MBH is present. The sensitivity of HWO will allow us, for the first time, to dynamically detect MBHs of mass $>10^{4.5} \, \rm M_\odot$ at the center of virtually any dwarf galaxy in the local Universe. Probing the presence of MBHs in local dwarf galaxies will allow us to study the scaling relations between black hole mass and host galaxy properties in detail. Generally, a scaling relation indicates how one quantity (e.g., the black hole mass) changes as a function of another (e.g., the host’s stellar mass). Coupled with data at higher redshift, this information will enable us to trace the time evolution of these scaling relations. 

As shown in Fig.~\ref{fig:fig1}, HWO, in its 8-m configuration, will be able to resolve the radius of influence and, thus, detect the presence of quiescent SMBHs of mass $>10^6 \,  \rm M_\odot$ in virtually any galaxy within $\sim$30 Mpc and reach down to a mass of $\sim10^5 \, \rm M_\odot$ in galaxies within $\sim$10 Mpc. This capability expands enormously our reach to discover the effect of the central black hole’s immense gravitational field on the core regions of galaxies.

\subsection{How Did the First Black Holes Form?}

Measuring the mass of central black holes in lower-mass, local galaxies will revolutionize our understanding of the mechanism that formed the first black holes around $z \sim 20$--$30$, i.e., the black hole ``seed'' population. Black hole seeds are the first population of black holes formed in the Universe (see, e.g., the reviews by \citealt{Woods_2019, Inayoshi_2020}), and they typically come in (at least) two flavors: light seeds (with mass $\sim10$--$1000 \, \rm M_\odot$) are formed from the remnants of Population III stars (see, e.g., \citealt{Barkana_Loeb_2001}), while heavy seeds (with masses $10^4$--$10^6 \, \rm M_\odot$) can be formed from the monolithic collapse of metal-free, atomic cooling halos \citep{Loeb_1994, Bromm_2003, Lodato_2006}.

Several works (e.g., the review by \citealt{Greene_2020}, and references therein) show that the scaling relations between the black hole mass and several galaxy properties, especially in the low-mass regime, fundamentally differ between heavy seeding and light seeding of the first black holes.

In summary, we report below the critical unknowns that this Science Case addresses. Note that the numbers quoted here are further justified and detailed in the upcoming Sections.

\subsubsection{Are Massive Black Holes in Local Dwarf Galaxies?}
\begin{itemize} 
    \item Utilize HWO's high-resolution UV and optical spectroscopy capabilities, in the range of 40--10 milli-arcseconds, to dynamically measure MBH masses down to $10^{4.5}$--$10^6 \, \rm M_\odot$. Within 10 Mpc, the 6.5-m configuration will allow reaching down to $10^6 \, \rm M_\odot$, while the 8-m configuration will open up the parameter space to $\sim10^{4.5} \, \rm M_\odot$.
    \item Focus on galaxies within a 10--30 Mpc radius to ensure probing the MBH's sphere of influence with HWO's exquisite angular resolution.
\end{itemize}

\subsubsection{Do Scaling Relations Between MBHs and Hosts in Local Dwarf Galaxies Evolve?}
\begin{itemize}
    \item Probe the scaling relations between MBH mass and various host galaxy properties, such as stellar mass and velocity dispersion. In particular, we aim at lowering the typical (dynamically measured) black hole mass from $10^{6.5}$--$10^6 \, \rm M_\odot$ (i.e., the current state of the art) down to $10^{4.5} \, \rm M_\odot$ in the optimal scenario. Similarly, in the most extended configuration, we aim to lower the typical stellar velocity dispersions from $80-50 \, \rm km \, s^{-1}$ to $30  \, \rm km \, s^{-1}$.

    \item Quantitatively measure if/how these relations differ (at a $>3\sigma$ level, where the standard scatter in the relations gives $\sigma$) in low-mass galaxies compared to more massive systems. Numerous theoretical studies suggest that this investigation will be vital to constraining the properties of the seed population of black holes.

    \item Provide insights into the early stages of galaxy formation and the role of MBHs in regulating star formation, thus co-evolving with the host.
\end{itemize}

\subsubsection{Did Heavy Black Hole Seeds Form?}

Probe the origins of MBHs by studying the low-mass end of their mass function, which is crucial for understanding the initial seed black holes formed in the early Universe. In particular:

\begin{itemize}
    \item The detection of a significantly undermassive or overmassive SMBH population at $>3\sigma$ level (compared to the intrinsic scatter for the local relation) would be a vital indication that light or heavy black hole seeding occurred (see, e.g., \citealt{Greene_2020}, \citealt{Pacucci_2023}, \citealt{Pacucci_2024}, \citealt{Pacucci_2025}). For example, detecting black holes with mass $>10^7 \, \rm M_\odot$ in hosts of $\sim10^9 \, \rm M_\odot$ in stars would support the overmassive hypothesis.
    
    \item The non-detection of any overmassive or undermassive populations would strongly indicate that, contrary to current theories, dwarf galaxies do not retain seeding information. This knowledge will nonetheless provide crucial information for models of co-evolution between galaxies and black holes, indicating that seeding information is deleted at high redshifts, even in the case of dwarf galaxies.
\end{itemize}

\section{Physical Parameters}

This Science Case aims to leverage HWO's capabilities to model nuclear stellar dynamics and determine the masses of quiescent MBHs in local dwarf galaxies. 

\subsection{Key Physical Parameters}

\subsubsection*{Black Hole Mass}

\begin{itemize}
    \item Measurements of MBH masses down to $\sim10^{4.5} \, \rm M_\odot$ (in galaxies of $\sim10^7 \, \rm M_\odot$ in stars, which are at the peak of the dwarf galaxy distribution in the local Universe) are needed to constrain seeding models, and confirm (or exclude at $>3\sigma$) the occurrence of light seeding.

    \item Determine the mass distribution of MBHs in local dwarf galaxies within 10--30 Mpc distance in a volume-limited sample, with a typical uncertainty of a factor $\sim2$--$3$.
\end{itemize}

\subsubsection*{Stellar Velocity Dispersion}
\begin{itemize}
\item Measure the velocity dispersion of stars in the central regions of galaxies to infer the presence, and estimate the mass, of MBHs, down to $30  \, \rm km \, s^{-1}$. This represents a significant breakthrough compared to the current state of the art of $80-50  \, \rm km \, s^{-1}$.

\item Obtain spatially resolved kinematic maps to study the influence of MBHs on the surrounding stellar populations.
\end{itemize}

\subsection{How Many Galaxies Can We Probe?}

\begin{figure*}[ht!]
    \centering
    \includegraphics[width=0.75\textwidth]{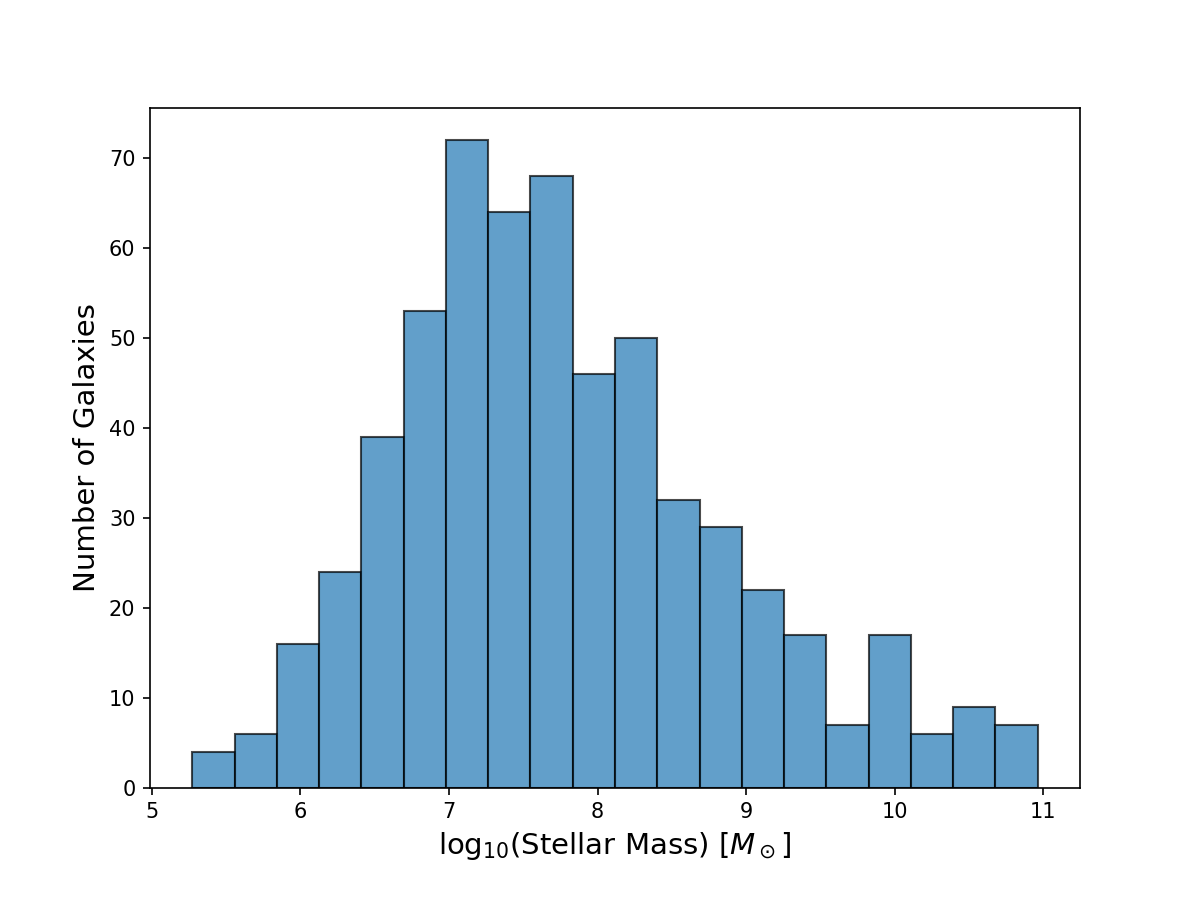}
    \caption{Mass distribution of galaxies within 10 Mpc. The peak of the distribution occurs for a stellar mass of $10^{7}$--$10^{8} \, \rm M_\odot$. There is a total of $\sim600$ galaxies within 10 Mpc. Figure created from the 50 Mpc Catalog \citep{Ohlson_2024}}
    \label{fig:fig3}
\end{figure*}

As detailed above, depending on its configuration, HWO can detect virtually any MBH hosted in local galaxies within a distance of $10-30$ Mpc. Detailed galaxy catalogs of the local Universe exist and will be fundamental to determining the targets of this search. For example, the ``50 Mpc Catalog'' reports 15,000+ galaxies within 50 Mpc \citep{Ohlson_2024} and $\sim$600 within 10 Mpc, as displayed in Fig.~\ref{fig:fig3}.

\subsection{Observational State of the Art and Science Targets}

To determine robust but, at the same time, revolutionizing goals for the present HWO Science Case, it is essential to investigate what the current state of the art is. From \citet{Kormendy_2013}, we can see that the black hole mass to stellar mass relation for quiescent galaxies is well sampled down to $\sim10^{6.5}$--$10^6 \, \rm M_\odot$ in black hole mass, which corresponds to $80-50 \, \rm km \, s^{-1}$ in stellar velocity dispersion (see also \citealt{Bennert_2021}). The region of parameter space for black hole masses $<10^6 \, \rm M_\odot$ is scarcely explored. HWO will allow the detection of dynamically quiescent black holes of such small mass at distances $\sim10$ Mpc.

Hence, we assemble the following Table \ref{tab:performance} to indicate what black hole masses and stellar velocity dispersions HWO would need to probe to reach incremental, substantial, and major progress, respectively, as well as the current state of the art. These values represent the black hole masses we expect in hosts of mass $10^9$, $10^8$, $10^7$, and $<10^7 \, \rm M_\odot$. The stellar mass of reference for these calculations is $10^{7.5}\, \rm M_\odot$, representing the peak in the stellar mass distribution of galaxies within 10 Mpc \citep{Ohlson_2024}.

\begin{table*}[ht!]
    \centering
    \caption[Performance Goals]{Black hole mass and velocity dispersion benchmarks for HWO configurations.}
    \label{tab:performance}
    \begin{tabular}{lcccc}
        \noalign{\smallskip}
        \hline
        \noalign{\smallskip}
        {Physical Parameter} & {State of the Art} & {Incremental Progress} & {Substantial Progress} & {Major Progress} \\
        \noalign{\smallskip}
        \hline
        \noalign{\smallskip}
        MBH Mass & $10^{6}$--$10^{6.5} \, \rm M_\odot$ & $10^{5.5}$--$10^{6} \, \rm M_\odot$ & $10^{5} \, \rm M_\odot$ & $<10^{5} \, \rm M_\odot$ \\
        Velocity Dispersion (Stars) & $50-80 \, \rm km \, s^{-1}$ & $40-50\, \rm km \, s^{-1}$ & $30 \, \rm km \, s^{-1}$ & $<30 \, \rm km \, s^{-1}$ \\
        \noalign{\smallskip}
        \hline
    \end{tabular}
\end{table*}

\begin{figure*}[ht!]
    \centering
    \includegraphics[width=0.75\textwidth]{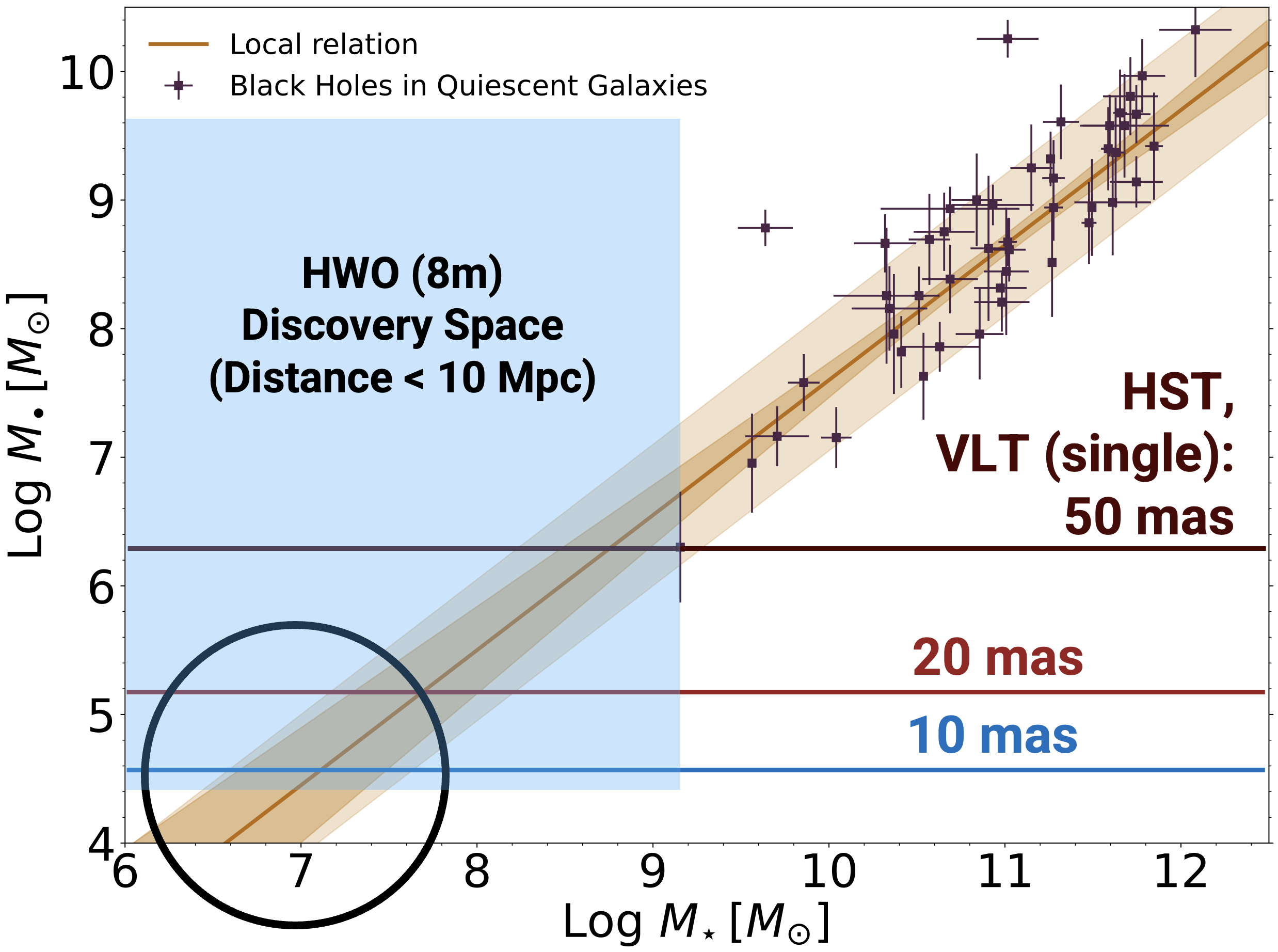}
    \caption{Discovery space opened up by HWO in its 8-m configuration for black holes up to 10 Mpc of distance. Data on quiescent galaxies from \cite{Bennert_2021}. We also display representative sensitivities to black hole masses for dwarf galaxies within 10 Mpc. The goal is to discriminate whether MBHs in local dwarf galaxies (whose location in the plot is highlighted by a black circle) follow the standard scaling relations between black hole mass and the host's stellar mass (or, similarly, with the stellar velocity dispersion of the host).}
    \label{fig:fig5}
\end{figure*}

The corresponding HWO discovery space is shown in Fig.~\ref{fig:fig5} for the 8-m configuration, within 10 Mpc of distance.

We also display the sensitivities to black hole mass that can be reached with a 20-mas and 10-mas configuration for HWO. These are compared to the state-of-the-art angular resolution, i.e., HST and VLT (in single-telescope operation), which are of the order of 50 mas.

\section{Description of Observations}

To achieve the science objectives outlined, HWO will conduct a series of observations designed to dynamically detect and study quiescent MBHs in local dwarf galaxies. These observations will leverage HWO's advanced capabilities in high-resolution spectroscopy and imaging.

\subsection{Target Dataset}

We propose UV and optical multi-object spectroscopy (MOS) of a volume-limited sample of $\sim100$ dwarf galaxies within 10--30 Mpc, primarily selected from the 50 Mpc Catalog \citep{Ohlson_2024}, which includes $\sim15{,}000$ galaxies. This catalog provides stellar mass estimates based on integrated photometry. However, precise stellar velocity dispersions are currently sparse or unavailable for most low-mass systems in this volume.

Future datasets from missions such as \textit{Roman}, \textit{Euclid}, and \textit{UVEX} could provide complementary rest-frame optical and UV photometry to refine stellar mass estimates via SED fitting, and detect nuclear point sources indicative of low-level accretion activity. These missions may also help identify cleaner targets by excluding AGNs, allowing a focus on dynamically quiescent systems.

\subsection{Selection Criteria}

Target galaxies will be selected to span a range of stellar masses between $10^6$ and $10^9 \, \rm M_\odot$, with priority given to systems where HWO can resolve the gravitational sphere of influence for expected black hole masses in the range of $10^{4.5}$ to $10^6 \, \rm M_\odot$. A minimum surface brightness threshold and isolation criteria may also be applied to ensure robust measurements of the stellar kinematics. Targets with suitable inclination angles and minimal dust attenuation in the nucleus will be preferred to optimize the extraction of velocity dispersion profiles.

\subsection{Sample Size Justification}

A sample of approximately 100 galaxies strikes a balance between feasibility and statistical significance. Assuming a Gaussian distribution of scatter around the MBH--host scaling relation, a sample of 100 objects enables constraints at the $\sim10\%$ level on the mean relation. This sample size is also sufficient to subdivide the population into bins by stellar mass (e.g., $10^6$--$10^7$, $10^7$--$10^8$, $10^8$--$10^9 \, \rm M_\odot$, etc.) and compare scaling relations or MBH detection fractions between them. We note that larger samples (e.g., 200+ objects) may be desirable for exploring second-order correlations and could be considered in follow-up phases.

\subsection{Observational Requirements}

The calculation on the required angular resolution, expressed in milli-arcseconds, is derived from the expression of the sphere of gravitational influence for a MBH at 10 Mpc: $R_{\rm soi} = GM_\bullet/\sigma^2$, where $G$ is the gravitational constant, $M_\bullet$ is the black hole mass, and $\sigma$ is the velocity dispersion.

At this distance, the 6.5-m configuration of HWO will be able to resolve the radius of gravitational influence for black holes down to $10^6 \, \rm M_\odot$. In the UV-Vis range, the wavelength range considered is 100--1000 nm. The velocity dispersion $\sigma$ is assumed to be the dispersion of the stellar component, calculated from the \citet{Kormendy_2013} relation (see Table~\ref{tab:obsreq}).

\begin{table*}[ht!]
    \centering
     \caption[Observation Requirements]{Angular resolution requirements for detecting quiescent MBHs in local dwarf galaxies.}
    \label{tab:obsreq}
    \begin{tabular}{lcccc}
        \noalign{\smallskip}
        \hline
        \noalign{\smallskip}
        {Observation} & {State of the Art} & {Incremental Progress} & {Substantial Progress} & {Major Progress} \\
        \noalign{\smallskip}
        \hline
        \noalign{\smallskip}
        Type & UV-Vis MOS (Ca~II triplet) & UV-Vis MOS & UV-Vis MOS & UV-Vis MOS \\
        Resolution & 50 mas & 44 mas & 20 mas & 10 mas \\
        \noalign{\smallskip}
        \hline
    \end{tabular}
\end{table*}

No significant restrictions on the Field of View (FoV) are required for these observations, as most of the spheres of influence are significantly smaller than the FoV of any planned detector. A standard field of view such as $1' \times 1'$ will be enough to image the entire region of interest, together with high PSF stability. In fact, given the extremely tight constraints on PSF stability for the primary science by HWO, we can safely assume that the PSF requirements for this science case are well-defined and well-characterized. Compared to HST, the PSF will be more stable; additionally, it will be more stable even compared to 30-m ground-based telescopes at wavelengths shorter than $\sim2~\mu$m.

Standard optical absorption features, such as the Ca~II triplet lines (around 850 nm), will be used to measure the stellar kinematics. These lines are optimal for kinematic measurements because they are typically uncontaminated by the ISM and well separated. For dwarf galaxies (i.e., a velocity dispersion of $<30 \, \rm km \, s^{-1}$), a spectral resolution of $\sim8000$ will be sufficient for these measurements.

\subsection{Why HWO Is Needed}

The unique combination of HWO's space-based, diffraction-limited imaging and high-resolution spectroscopy in the UV-optical makes it the only facility capable of delivering this science:

\begin{itemize}
    \item Ground-based ELTs, while powerful, depend on adaptive optics systems that require nearby guide stars and perform sub-optimally in the optical regime. This constraint limits their usable target pool and makes them impractical for the low-surface-brightness dwarf galaxies considered here.
    \item ELTs are also unlikely to succeed in measuring MBH masses in active galaxies, as nuclear emission overwhelms the stellar continuum needed for kinematic measurements.
    \item HWO offers stable PSFs, full sky access, and high-resolution spectroscopy in the required bands---all without atmospheric interference---making it uniquely capable of resolving spheres of influence as small as a few parsecs at distances of 10--30 Mpc.
\end{itemize}

{\bf Acknowledgements.} F.P. acknowledges support from a Clay Fellowship administered by the Smithsonian Astrophysical Observatory. This work was also supported by the Black Hole Initiative at Harvard University, which is funded by grants from the John Templeton Foundation and the Gordon and Betty Moore Foundation. 

\bibliography{author.bib}

\end{document}